\documentclass[a4paper,10pt]{article}

\usepackage{amssymb}
\usepackage{amsmath}
\usepackage{subfigure}
\usepackage{slashed}

\allowdisplaybreaks

\usepackage{soul}

\usepackage{graphicx}
\usepackage{amsthm}
\usepackage{latexsym}
\usepackage[dvips]{epsfig}

\usepackage{wasysym}
\usepackage{mathrsfs}
\usepackage{eufrak}
\usepackage{bm}
\usepackage{authblk}
\usepackage{slashed}
\usepackage{yhmath} 
\usepackage{stmaryrd}

\theoremstyle{plain}
\newtheorem{proposition}{Proposition}
\newtheorem{lemma}{Lemma}
\newtheorem{theorem}{Theorem}

\newtheorem{corollary}{Corollary}

\newtheorem{definition}{Definition}
\newtheorem{remark}{Remark}

\def\bma{{\bm a}}
\def\bmb{{\bm b}}
\def\bmc{{\bm c}}

\def\bme{{\bm e}}

\def\bmg{{\bm g}}
\def\bmh{{\bm h}}

\def\bmzero{{\bm 0}}
\def\bmone{{\bm 1}}

\def\bmA{{\bm A}}
\def\bmB{{\bm B}}
\def\bmC{{\bm C}}
\def\bmD{{\bm D}}

\def\bmT{{\bm T}}

\def\bmeta{{\bm \eta}}

\def\bmxi{{\bm \xi}}

\def\bmmu{{\bm \mu}}

\def\bmpi{{\bm \pi}}

\def\bmsigma{{\bm \sigma}}

\def\bmnabla{{\bm \nabla}}

\newcounter{mnotecount}

\newcommand{\mnotex}[1]
{\protect{\stepcounter{mnotecount}}$^{\mbox{\footnotesize $\bullet$\themnotecount}}$ 
\marginpar{
\raggedright\tiny\em
$\!\!\!\!\!\!\,\bullet$\themnotecount: #1} }

\usepackage[utf8]{inputenc}
\usepackage[english]{babel}
 
\usepackage{multicol}
\usepackage{color}
 
\usepackage{comment}
 
\begin{document}

\title{\textbf{ Staticity and regularity for zero rest-mass fields near spatial infinity on flat spacetime}}

\author[1]{E. Gasper\'in\footnote{E-mail address:{\tt
      edgar.gasperi@tecnico.ulisboa.pt}}} \author[,2]{J. A. Valiente
  Kroon \footnote{E-mail address:{\tt j.a.valiente-kroon@qmul.ac.uk}}}
  \affil[1] {CENTRA, Departamento de F\'isica, Instituto Superior
  T\'ecnico IST, Universidade de Lisboa UL, Avenida Rovisco Pais 1,
  1049 Lisboa, Portugal}
\affil[1a]{Institut  de  Math\'ematiques  de  Bourgogne  (IMB),  UMR  5584,  CNRS, Universit\'e  de  Bourgogne  Franche-Comt\'e,  F-21000  Dijon,  France.}
\affil[2]{School of Mathematical Sciences,
  Queen Mary, University of London, Mile End Road, London E1 4NS,
  United Kingdom.}

\maketitle

\begin{abstract}
Linear zero-rest-mass fields generically develop
logarithmic singularities at the critical sets where spatial infinity
meets null infinity. Friedrich's representation of spatial infinity is
ideally suited to study this phenomenon. These logarithmic singularities are an
obstruction to the smoothness of the zero-rest-mass field at null
infinity and, in particular, to peeling. In the case of the spin-2
field it has been shown that these logarithmic singularities can be
precluded if the initial data for the field satisfies a certain
regularity condition involving the vanishing, at spatial infinity, of a certain spinor (the
linearised Cotton spinor) and its totally symmetrised derivatives. In
this article we investigate the relation between this regularity
condition and the staticity of the spin-2 field. It is shown that
while any static spin-2 field satisfies the regularity condition,
not every solution satisfying the regularity condition is static.
This result is
in contrast  with what happens in the case of General Relativity where
staticity in a neighbourhood of spatial infinity and the smoothness of the
field at future and past null infinities are much more closely
related.
\end{abstract}

\textbf{Keywords:} Conformal methods, 
spinors, staticity, cylinder at spatial infinity, regularity.

\medskip
\textbf{PACS:} 04.20.Ex, 04.20.Ha, 04.20.Gz

\section{Introduction}
Obtaining a satisfactory understanding of the consequences of the
degeneracy of the conformal structure of spacetime at spatial infinity
is one of the key open problems in the mathematical theory of isolated
systems in General Relativity ---the so-called \emph{problem of spatial
  infinity}. A programme to systematically investigate this problem
was initiated with the seminal work of Friedrich in \cite{Fri98a}. The
key idea of this work is the development of a representation of
spatial infinity (\emph{the cylinder at spatial infinity}) which
allows the formulation of \emph{a regular Cauchy problem in a neighbourhood
of spatial infinity} for the \emph{conformal Einstein field
equations}. This framework allows to show that, unless the initial
data is fine-tuned, the solutions to the conformal Einstein field
equations (and in particular the components of the rescaled Weyl
tensor) develop logarithmic singularities at the \emph{critical sets} where
null infinity meets spatial infinity. In the particular case of \emph{time-symmetric
initial data sets} for the Einstein field equations which admit a point
compactification at infinity for which the resulting \emph{conformal metric
is analytic} then its possible to show that a certain subset of the
logarithmic singularities can be avoided if the conformal metric $\bmh$
satisfies the (conformally invariant) condition
\begin{equation}
D_{\{i_p \cdots i_1} b_{jk\}}(i)=0, \qquad p=0,\; 1,\; 2,\ldots ,
\label{GRRegularityCondition}
\end{equation}
where $b_{ij}$ denotes the Cotton-Bach tensor of the metric $\bmh$ and
${}_{\{\cdots \}}$ denotes the operation of computing the symmetric
tracefree part
---in particular, if $\bmh$ is conformally flat then $
b_{jk}=0$. Observe that this condition is imposed at the point at
infinity of the asymptotic end. Accordingly, it is a truly asymptotic
condition. The
regularity condition \eqref{GRRegularityCondition} has also arisen in
slightly different contexts and is satisfied by
static initial data sets ---see \cite{Fri88,Bei91b}. Subsequent
analysis has shown that although condition
\eqref{GRRegularityCondition} is a necessary condition for avoiding
logarithmic singularities at the critical sets, it is by no means
sufficient ---see \cite{Val04a,Val04d}. Static solutions to the
Einstein field equations have been shown to be logarithmic
free at the critical points of Friedrich's representation of spatial
infinity \cite{Fri04}. Moreover, the analysis in \cite{Val10,Val11} strongly
suggests the conjecture that, among the class of time symmetric
initial data sets, \emph{only those which are static in a neighbourhood of
infinity will give rise to developments which are free of logarithmic
singularities at the critical sets} ---see also \cite{Fri13}. The gluing techniques developed in
\cite{CorSch06,ChrDel03} allow the construction of large classes of
initial data sets with this property. The interested reader is
directed to \cite{Fri18} for further discussion of this extensive
topic ---see also \cite{CFEBook}, Chapter 21.

\medskip
Generically, linearised fields propagating on the Minkowski spacetime also
develop logarithmic singularities at the critical sets ---see
e.g. \cite{Val03a}. In particular, for the massless spin-2 field,
there is an analogue of condition \eqref{GRRegularityCondition} which,
in this case, \emph{precludes the development of all logarithmic
  singularities} at least for a large class of initial data regular at
infinity. This condition is expressed in terms of the totally symmetric
\emph{linearised Cotton-Bach spinor} $B_{ABCD}$ as
\begin{equation}
D_{(A_pB_p} \cdots D_{A_1B_1} B_{ABCD)}(i)=0, \qquad
p=0,\,1,\,2,\ldots,
\label{LinearisedRegularityCondition}
\end{equation}
where $D_{AB}$ denotes the spinorial counterpart of the 3-dimensional
Levi-Civita connection $D_i$. The spinor $B_{ABCD}$ is essentially the
curl of the physical spin-2 field. 

Given the strong relation between the regularity condition
\eqref{GRRegularityCondition} and the staticity of the associated
spacetime, it is natural to ask whether there exists a connection
between condition \eqref{LinearisedRegularityCondition} and static
spin-2 fields ---i.e. spin-2 fields whose Lie derivative with respect
to the static Killing vector of the Minkowski spacetime vanish. This
is the question addressed in this article. Our
analysis shows that \emph{while the regularity condition
\eqref{LinearisedRegularityCondition} is satisfied by any such static
spin-2 fields, not every spin-2 field arising from initial data for
which condition \eqref{LinearisedRegularityCondition} holds is
static}. More precisely, the spin-2 field will be static if and only
if 
\[
B_{ABCD}=0 \qquad \mbox{on the initial hypersurface}
\]
---see Theorem \ref{Theorem:Main}.  In particular, $B_{ABCD}$ and its derivatives to all orders vanish at
$i$.  Accordingly, the requirement of smoothness at null infinity
is a far less stringent requirement for the linearised field than for
the (non-linear) gravitational field. This observation should serve as
a caution towards extracting conclusions about asymptotics of the
gravitational field from an analysis of of the linearised equations. 

\subsection*{Outline of the article}
Section \ref{CylinderAtspatialInfinity} provides a brief overview of
Friedrich's representation of spatial infinity in terms of a
cylinder. Section \ref{RegularityCondition} provides a discussion of
the properties of the spin-2 field which are relevant for the analysis
in this article. Section \ref{SpaceSpinorFormalism} contains a brief
overview of the space-spinor formalism used to analyse the spin-2
field. Section \ref{Section:Staticity} provides a study of
static spin-2 fields.

\subsection*{Notations and Conventions}
The signature convention for (Lorentzian) spacetime metrics will be $
 (+,-,-,-)$.  In the rest of this article $\{_a , _b , _c , . . .\}$
 denote abstract tensor indices and $\{_\bma , _\bmb , _\bmc , . . .\}$
 will be used as spacetime frame indices taking the values ${ 0,
   . . . , 3 }$.  In this way, given a basis $\{\bme_{\bma}\}$ a
 generic tensor is denoted by $T_{ab}$ while its components in the
 given basis are denoted by $T_{\bma \bmb}\equiv
 T_{ab}\bme_{\bma}{}^{a}\bme_{\bmb}{}^{b}$. Greek indices $\{_\mu ,
 _\nu , _\lambda, \ldots  \}$ are used to denote spacetime coordinates
 while $\{_\alpha , _\beta , \ldots\}$ play the role of spatial coordinate
 indices. The core of the  analysis
 makes use of spinors. For this, the notation and
 conventions of Penrose \& Rindler \cite{PenRin84} will be followed.
 In particular, capital Latin indices $\{ _A , _B , _C , . . .\}$ will
 denote abstract spinor indices while boldface capital Latin indices
 $\{ _\bmA , _\bmB , _\bmC , . . .\}$ will denote frame spinorial
 indices with respect to a specified spin dyad ${ \{\delta_\bmA{}^{A}
   \} }.$
The  conventions for the curvature tensors are fixed by the relation
\[
(\nabla_a \nabla_b -\nabla_b \nabla_a) v^c = R^c{}_{dab} v^d.
\]

\section{Spatial infinity in the Minkowski spacetime}
\label{CylinderAtspatialInfinity}

In this section we provide a quick overview of Friedrich's
representation of spatial infinity for the Minkowski spacetime. The
reader interested in further details is referred to \cite{Fri98a,Fri03b,Val03a,CFEBook,MagVal21}.

\medskip
Let $(\mathbb{R}^4, \tilde{\bmeta})$ denote the Minkowski spacetime
and consider Cartesian coordinates $(\tilde{x}^{\mu})=
(\tilde{t},\tilde{x}^\alpha)$ in which the Minkowski line element reads
 \begin{equation}\label{PhysicalMinkowskiMetricCartesianCoords}
\tilde{\bmeta}=\eta_{\mu\nu}\mathbf{d}\tilde{x}^{\mu}\otimes\mathbf{d}\tilde{x}^{\nu},
\qquad \eta_{\mu\nu}\equiv\text{diag}(1,-1,-1,-1).
\end{equation}
  Let
 $(\tilde{t},\tilde{\rho},\theta^{\mathcal{A}})$ with $\mathcal{A}=2,\,3$ denote spherical
 polar coordinates defined by $\tilde{\rho}^2\equiv 
 \delta_{\alpha\beta}\tilde{x}^{\alpha}\tilde{x}^{\beta}$ where
 $\delta_{\alpha\beta}\equiv\text{diag(1,1,1)}$, and $(\theta^{\mathcal{A}})$ represents an
 arbitrary choice of coordinates on $\mathbb{S}^2$. In these
 coordinates the metric $\tilde{\bmeta}$ is given by
\begin{equation*}
\tilde{\bmeta}=\mathbf{d}\tilde{t}\otimes\mathbf{d}\tilde{t}
-\mathbf{d}\tilde{\rho}\otimes \mathbf{d}\tilde{\rho}-\tilde{\rho}^2
\mathbf{\bm\sigma},
\end{equation*}
with $\tilde{t}\in(-\infty,\infty)$, $\tilde{\rho}\in [0,\infty)$ and
  $\bm\sigma$ represents the standard round metric on $\mathbb{S}^2$.

\subsection{The basic formalism}
In the sequel we restrict our discussion to the region
\[
\mathcal{D}\equiv \{ \tilde{x}^\mu \;|\; \eta_{\mu\nu}
\tilde{x}^\mu \tilde{x}^\nu<0 \}
\]
which contains the asymptotic region of the Minkowski spacetime around
spatial infinity. On $\mathcal{D}$ introduce \emph{inversion coordinates} $(x^{\mu})=(t,x^\alpha)$
  defined by
 \[
 x^{\mu}=-{\tilde{x}^{\mu}}/{\tilde{X}^2}, \qquad
  \tilde{X}^2 \equiv
  \eta_{\mu\nu}\tilde{x}^{\mu}\tilde{x}^{\nu}.
  \]
The inverse transformation is given by 
 \[\tilde{x}^{\mu}=-x^{\mu}/X^2, 
\qquad
 X^2=\eta_{\mu\nu}x^{\mu}x^{\nu},
\] 
with 
\[
X^2=1/\tilde{X}^2.
\]

\subsubsection{The standard representation of spatial infinity}
 Rewriting the physical Minkowski metric $\tilde{\bmeta}$ in these
 coordinates leads to a conformal representation with conformal metric
 given by 
\begin{equation}
\label{InverseMinkowskiMetricDef}
\bar{\bmeta}=\Xi^2 \hspace{0.5mm}\tilde{\bmeta},
\end{equation}
where 
\[
\bar{\bmeta}=\eta_{\mu\nu}\mathbf{d}x^{\mu}\otimes
\mathbf{d}x^{\nu}, \qquad \Xi =X^2.  
\]
Observe that in this representation, the conformal metric  $\bar{\bmeta}$ is, locally the
Minkowski metric ---in particular, the origin of
$(\mathbb{R}^4,\bar{\bmeta})$ corresponds to the the spatial infinity
$i^0$ of the Minkowski spacetime $(\mathbb{R}^4,\tilde{\bmeta})$. 
Introducing  \emph{unphysical
  polar coordinates} $(t,\rho,\theta^{\mathcal{A}})$ where
$\rho^2\equiv \delta_{\alpha\beta}x^{\alpha}x^{\beta}$,
the metric $\bar{\bmeta}$ and conformal factor
$\Xi$ read
\begin{equation}
\label{InverseMinkowskiUnphysicaltrhocoords}
\bar{\bmeta}=\mathbf{d}t\otimes\mathbf{d}t -\mathbf{d}\rho\otimes
\mathbf{d}\rho-\rho^2 \mathbf{\bm\sigma}, \qquad \Xi=t^2-\rho^2,
\end{equation}
with $t\in(-\infty,\infty)$ and $\rho\in (0,\infty)$. Unwrapping the
above definitions, shows that relation between the physical and
unphysical polar coordinates is given by
\begin{equation}
\label{UnphysicalToPhysicaltrho}
\tilde{t}=-\frac{t}{t^2-\rho^2}, \qquad \tilde{\rho}=
\frac{\rho}{t^2-\rho^2}.
\end{equation}

\subsubsection{Friedrich's representation}
In order to introduce Friedrich's representation of spatial infinity
(the so-called \emph{F-gauge}),
one considers a new time coordinate $\tau$ defined via the relation 
\[
t=\rho\tau.
\]
In terms of this new coordinate the conformal metric $\bar{\bmeta}$ reads
\begin{equation*}
\bar{\bmeta}=\rho^2 \mathbf{d}\tau\otimes \mathbf{d}\tau
-(1-\tau^2)\mathbf{d}\rho \otimes \mathbf{d}\rho + \rho\tau
(\mathbf{d}\rho\otimes \mathbf{d}\tau + \mathbf{d}\tau \otimes
\mathbf{d}\rho) - \rho^2 \bmsigma.
\end{equation*}
The above line element suggests a new conformal representation of the
Minkowski spacetime with conformal metric given by
\begin{equation}
\label{MetricCylinderToIversedMinkowski}
\bmeta \equiv \frac{1}{\rho^2} \bar{\bmeta}.
\end{equation}
Writing the metric $\bmeta$ explicitly one finds that 
\begin{equation}
\bmeta = \mathbf{d}\tau\otimes \mathbf{d}\tau
-\frac{(1-\tau^2)}{\rho}\mathbf{d}\rho \otimes \mathbf{d}\rho +
\frac{\tau}{\rho}(\mathbf{d}\rho\otimes \mathbf{d}\tau +
\mathbf{d}\tau \otimes\mathbf{d}\rho) - \bmsigma,
\label{FMinkowski}
\end{equation}
which is clearly is singular at $\rho=0$. Notice, however, that its contravariant version
\[
\bmeta^{\sharp}=(1-\tau^2)\bm\partial_\tau\otimes \bm\partial_{\tau}
+ \rho\tau(\bm\partial_\tau\otimes \bm\partial_\rho
+\bm\partial_\rho\otimes \bm\partial_\tau)
-\rho^2\bm\partial_\rho\otimes \bm\partial_\rho - \bm\sigma^{\sharp},
\]
is regular ---see \cite{MagVal21} for further discussion on this
peculiarity of Friedrich's representation of spatial
infinity. Associated to the above contravariant line element one has a 
null frame $\{\bme_{\bmA\bmA'}\}$ such that 
\[
\bmeta^\sharp=\epsilon^{\bmA
  \bmB}\epsilon^{\bmA'\bmB'}\bme_{\bmA\bmA'}\otimes \bme_{\bmB\bmB'},
\]
with
\begin{subequations}
\begin{eqnarray}
&&\bme_{\bm0\bm0'}=\frac{1}{\sqrt{2}}\big((1-\tau)\bm\partial_{\tau} +
  \rho\bm\partial_{\rho}\big), \label{Fframe00}\\
&&\bme_{\bm1\bm1'}=\frac{1}{\sqrt{2}}\big((1+\tau)\bm\partial_{\tau} -
  \rho\bm\partial_{\rho}\big), \label{Fframe11}\\
&&\bme_{\bm0\bm1'}=\frac{1}{\sqrt{2}}\bm\partial_{+}, \\
&& \bme_{\bm1\bm0'}=\frac{1}{\sqrt{2}}\bm\partial_{-}, \label{Fframe0110}
\end{eqnarray}
\end{subequations}
where$\{ \bm\partial_{+},\bm\partial_{-}\}$ is a complex null frame on
\[
\mathcal{Q}_{\tau,\rho}\equiv \{p\in \mathcal{M} \;|\; \tau= const,\;\;
\rho=const\} \approx \mathbb{S}^2
\]
satisfying 
\[
[\bm\partial_{\tau},\bm\partial_{\pm}]=0, \qquad
[\bm\partial_{\rho},\bm\partial_{\pm}]=0.
\]
As discussed in the Appendix of \cite{GasVal20}, the use of a specific
choice of coordinates $(\theta^{\mathcal{A}})$ for
$\mathbb{S}^2$ is not required in the subsequent discussion.
Another way to deal with the the deficiencies of spherical
coordinates is to lift  the structures to a suitable \emph{bundle space}.
This approach is
briefly discussed in Section \ref{Section:FibreSpace}.

\subsubsection{The structure of the conformal boundary}
In the conformal representation introduced in the previous section
spatial infinity corresponds to the set
of points with $\rho=0$ and has the topology of
$\mathbb{R}\times\mathbb{S}^2$ ---hence the name of the \emph{cylinder
  at spatial infinity}. The rescaling relating the physical
Minkowski spacetime $(\mathbb{R}^4, \tilde{\bmeta})$ and the cylinder
conformal representation $(\mathcal{M},\bmeta)$ is given by
\begin{equation}
\label{ThetaInUnphysicalCoords}
\bmeta=\Theta^2 \tilde{\bmeta}, \qquad \Theta\equiv \rho(1-\tau^2),
\end{equation}
and
\[
\mathcal{M} \equiv \{ p \in \mathbb{R}^4 \; \rvert \; -1 \leq \tau(p)
\leq 1 , \; \; \rho(p)\geq 0\}.
\]
This representation is closely related to the properties of conformal
geodesics ---see e.g. \cite{MagVal21}. However, this feature will not
be exploited in the sequel. The physical Minkowski spacetime corresponds to the region
\[
\tilde{\mathcal{M}} \equiv \{ p \in \mathcal{M} \; \rvert \; -1<\tau(p)<1 , \; \;\rho(p)>0 \},
\]
while future and past null infinity are
located at
\begin{eqnarray*}
 && \mathscr{I}^{+} \equiv \{ p \in \mathcal{M} \; \rvert\; \tau(p) =1 \}, \\
&& \mathscr{I}^{-} \equiv \{ p \in \mathcal{M} \; \rvert \; 
 \tau(p) =-1\},
 \end{eqnarray*}
and spatial infinity is ``blown-up'' to a set $\mathcal{I}
 \approx \mathbb{R} \times \mathbb{S}^2$ given by 
\[
\mathcal{I} \equiv \{ p \in \mathcal{M} \; \rvert   \;\;  |\tau(p)|<1, \; \rho(p)=0\}, 
\qquad I^{0} \equiv \{ p \in \mathcal{M}\; \rvert \;
  \tau(p)=0, \; \rho(p)=0\}.
\]
Moreover, one identifies the critical sets
\begin{eqnarray*}
&& \mathcal{I}^{+} \equiv \{ p\in \mathcal{M} \; \rvert \; \tau(p)=1, \; \rho(p)=0
  \}, \\ 
&& \mathcal{I}^{-} \equiv \{p \in \mathcal{M}\; \rvert \; \tau(p)=-1, \; \rho(p)=0\},
\end{eqnarray*}
corresponding to the sets where spatial infinity ``touches'' null
infinity. Additionally, let
\[
\tilde{\mathcal{S}}_\star=\{ p\in \mathbb{R}^{4} \;| \;\tilde{t}(p)=0 \}, \qquad
\mathcal{S}_\star=\{ p\in \mathcal{M} \; | \;\tau(p)=0 \},
\]
describing the time symmetric hypersurface of the Minkowski
spacetime. Observe that $\tilde{\mathcal{S}}_\star$ and the interior
of $\mathcal{S}_\star$ coincide as sets of points. The region where $\mathcal{S}_\star$ intersects $\mathcal{I}$ will
be denoted as $\mathcal{I}^0$.

\subsection{The fibre space}
\label{Section:FibreSpace}
As it is well known, the vectors $\bme_{\bmzero\bmone'}$ and
$\bme_{\bmone\bmzero'}$ cannot define non-vanishing smooth vector fields everywhere over a
manifold with the topology of $\mathbb{S}^2$. In order to deal with
this technical difficulty, the equations are lifted to a 5-dimensional
submanifold of the bundle of normalised spin frames. The details of
this construction and its particularisation to the case of Minkowski
spacetime has been detailed in \cite{Fri98a}. In the following
paragraph we provide a brief overview of this construction. 

\medskip
Rotations of the form $\bme_{\bmzero\bmone'}\mapsto
e^{i\vartheta}\bme_{\bmzero\bmone'}$,
$\vartheta\in\mathbb{R}$, in the planes orthogonal to
$\bme_{\bmzero\bmzero'}$ and $\bme_{\bmone\bmone'}$ leave these
submanifolds invariant. Accordingly, it defines a subbundle with
structure group $U(1)$ which projects onto
$\mathcal{M}\setminus\mathcal{I}$. This projection is given by the
\emph{Hopf map} $SU(2)\rightarrow SU(2)/ U(1)\simeq
\mathbb{S}^2$. All the structures on $\mathcal{M}\setminus\mathcal{I}$
are then lifted to this subbundle. In an abuse of notation, we make
use of the same symbols to denote the original objects on
$\mathcal{M}\setminus\mathcal{I}$  and their lifted counterparts. On
the subbundle we consider coordinates $\tau$, $\rho$ and $s\in
SU(2)$. The lift is carried out in such a way that the lifted fields
$\bme_{\bmzero\bmzero'}$ and $\bme_{\bmone\bmone'}$ have the same
coordinate expressions as in equations
\eqref{Fframe00}-\eqref{Fframe11}. The coordinate $\rho$ is then
extended in a natural manner to include this value.  In terms of the
coordinates $(\tau,\rho,s)$ on the extended bundle, to be denoted again
by $\mathcal{M}$, the the lifted metric and conformal factor are
given, again, by \eqref{FMinkowski} and
\eqref{ThetaInUnphysicalCoords}. In this context, $\bmsigma$ denotes
the pull-back of the line element on $\mathbb{S}^2$ to
$SU(2)$. Accordingly, one has that
\[
\mathcal{M}\simeq [-1,1]\times [0,\infty]\times SU(2),
\]
while
\[
\mathcal{I}\simeq [-1,1]\times SU(2), \qquad \mathcal{I}^0 \simeq SU(2),
\qquad \mathcal{I}^\pm \simeq SU(2), \qquad \mathscr{I}^\pm \simeq
\mathbb{R} \times SU(2),
\]
are now considered as subsets of the subbundle $\mathcal{M}$. 

\medskip
In order to define vector fields on the $SU(2)$ part of the subbundle,
consider the basis
\[
u_1 =\frac{1}{2}
\left(
\begin{array}{cc}
0 & \mbox{i} \\
\mbox{i} & 0
\end{array}
\right), \qquad
u_2 =\frac{1}{2}
\left(
\begin{array}{cc}
0 & -1 \\
1 & 0
\end{array}
\right), \qquad
u_3 =\frac{1}{2}
\left(
\begin{array}{cc}
\mbox{i} & 0 \\
0 & -\mbox{i}
\end{array}
\right),
\]
of the Lie algebra of $SU(2)$ with commutator
$[u_i,u_j]=\epsilon_{ijk}u_k$. Denote by
$\mathbf{Z}_1,\,\mathbf{Z}_2,\, \mathbf{Z}_3$, the left invariant
vector fields generated by $u_1,\,u_2,\,u_3$ on the Lie group
$SU(2)$. In particular, $\mathbf{Z}_3$ is the vertical vector field
which generates the group $U(1)$ acting on the fibres of $\mathcal{M}$
and define $\mathbf{X} \equiv -2\mbox{i}\mathbf{Z}_3$. Finally, define
the complex conjugate vector fields
\[
\mathbf{X}_\pm \equiv - ( \mathbf{Z}_2 \pm \mbox{i} \mathbf{Z}_1),
\] 
and define
\[
\bme_{\bmzero\bmone'} = -\frac{1}{\sqrt{2}}\mathbf{X}_+, \qquad
\bme_{\bmone\bmzero'} = -\frac{1}{\sqrt{2}}\mathbf{X}_-.
\]
The construction outline above leads to vector fields 
\[
\bme_{\bmA\bmA'}, \qquad \mathbf{Z}_3 \qquad \mbox{on}\quad \mathcal{M}\setminus\mathcal{I}
\]
which extend smoothly to $\mathcal{I}$ and satisfy
\[
\bmeta(\mathbf{Z}_3,\cdot) =0, \qquad
\bmeta(\bme_{\bmA\bmA'},\bme_{\bmB,\bmB'})
=\epsilon_{\bmA\bmB}\epsilon_{\bmA'\bmB'} \qquad \mbox{on} \quad \mathcal{M}\setminus\mathcal{I}.
\]
Finally, the connection form induced on $\mathcal{M}\setminus\mathcal{I}$ by
the corresponding connection form on the bundle of normalised spin
coefficients defines the following non-vanishing connection
coefficients with respect to the frame $\{
\bme_{\bmA\bmA'} \}$:
\[
\Gamma_{\bmzero\bmzero'\bmzero\bmone} =
\Gamma_{\bmone\bmone'\bmzero\bmone} =-\frac{1}{2\sqrt{2}}.
\]
All the other connection coefficients vanish.

\section{The spin-2 field equation and the regularity condition}
\label{RegularityCondition}

The interest on the spin-2 equation stems from the fact that it can be
used to study the linearised gravitational field and as a model for
the Bianchi equations satisfied by the components of the Weyl tensor. In contrast with
more traditional approaches to linearised gravity where the key unknown is a metric perturbation, in this case, the linearised
gravitational field is encoded in a tensor representing the weak-field
Riemann curvature \cite{PenRin86, Val03a}.  If the weak-field limit of
the vacuum Einstein field equations are imposed, it
becomes a traceless tensor (hence with the same symmetries of the Weyl curvature
tensor) and can be succinctly described by a totally symmetric spinor
$\phi_{ABCD}$ ---see \cite{PenRin86}. The linearised Bianchi
equations provide the following spinorial field equations for the
field $\phi_{ABCD}$:
\begin{equation}\label{Spin2Equation}
\nabla_{A'}{}^{A}\phi_{ABCD}=0.
\end{equation}
Due to its conformal properties ---see Lemma
\ref{conf_invariance_spin2}--- the spin-2 equation
\eqref{Spin2Equation} can also be interpreted as a toy model for the
conformal Einstein field equations ---see \cite{Val03a}.  The central
point to be discussed in this section is that the spin-2 field
$\phi_{ABCD}$ propagating in $(\mathcal{M},\bmeta)$ has, in general, 
polyhomogeneous solutions. This was originally shown in \cite{Val03a}
in the language of fibre bundles and making use of the fibre space
discussed in Subsection \ref{Section:FibreSpace} ---\cite{Fri98a, Fri03b}.  A similar analysis of
the solutions that avoids the use of this extended bundle space was
given in \cite{GasVal20} for the calculation of the Newman-Penrose
constants. In this section we review part of these constructions and
recall the regularity condition found in \cite{Val03a} that controls
the appearance ---at the level of initial data---
of the logarithmic terms in a Taylor like expansion for
the components of $\phi_{ABCD}$ close to $i^0$.

\subsection{Component expressions}

Let $\{\epsilon_{\bmA}{}^A\}$ be a spinor dyad with
$\epsilon_{\bm0}{}^A=o^A$ and $\epsilon_{\bm1}{}^A=\iota^A$.  In terms
of this spin dyad, the spinor $\phi_{ABCD}$ is encoded in the
following five complex scalars
\begin{eqnarray*} 
 &\phi_{0} \equiv\phi_{ABCD}o^{A}o^{B}o^{C}o^{D}, \qquad 
  \phi_{1} \equiv\phi_{ABCD}o^{A}o^{B}o^{C}\iota^{D},  \qquad 
  \phi_{2} \equiv\phi_{ABCD}o^{A}o^{B}\iota^{C}\iota^{D},& \\ 
 & \phi_{3}  \equiv\phi_{ABCD}o^{A}\iota^{B}\iota^{C}\iota^{D},
   \qquad 
  \phi_{4} \equiv\phi_{ABCD}\iota^{A}\iota^{B}\iota^{C}\iota^{D},
\end{eqnarray*}
which have spin weight $2,\,1,\,0,\,-1,\,-2$ respectively. In the
following we consider coefficients $\phi_0,\,\ldots, \,\phi_4$ as
their lift to the fibre space introduced in Section
\ref{Section:FibreSpace}. A direct calculation shows that the spin-2  equation \eqref{Spin2Equation}
implies the system of evolution equations 
\begin{subequations}
\begin{align}
& (1+\tau)\bm\partial_{\tau}\phi_{0}-\rho\bm\partial_{\rho}\phi_{0} +
  \mathbf{X}_+\phi_{1}=-2\phi_{0}, \label{EvolutionGravEq0}\\ &
  \bm\partial_{\tau}\phi_{1} + \frac{1}{2} \mathbf{X}_-\phi_{0}+
  \frac{1}{2}\mathbf{X}_+\phi_{2}=-\phi_{1}, \label{EvolutionGravEq1}\\ &
  \bm\partial_{\tau}\phi_{2} +\frac{1}{2}\mathbf{X}_-\phi_{1}+\frac{1}{2}
\mathbf{X}_+ \phi_{3}=0,
 \label{EvolutionGravEq2}
  \\ &
  \bm\partial_{\tau}\phi_{3}+\frac{1}{2}\mathbf{X}_-\phi_{2}+\frac{1}{2}\mathbf{X}_+\phi_{4}
 =\phi_{3}, \label{EvolutionGravEq3}\\ &
  (1-\tau)\bm\partial_{\tau}\phi_{4}+ \rho
  \bm\partial_{\rho}\phi_{4}+ \mathbf{X}_- \phi_{3} = 2\phi_{4},
 \label{EvolutionGravEq4}
\end{align}
\end{subequations}
and constraint equations
\begin{subequations}
\begin{align}
&  \tau\bm\partial_{\tau}\phi_{1}-\rho\bm\partial_{\rho}\phi_{1} + \frac{1}{2}
 \mathbf{X}_+\phi_{2} - \frac{1}{2}\mathbf{X}_-\phi_{0} =0, \label{ConstraintGravEq5}
\\ &
  \tau\bm\partial_{\tau}\phi_{2}-\rho\bm\partial_{\rho}\phi_{2}
 + \frac{1}{2}\mathbf{X}_+ \phi_{3} - \frac{1}{2}\mathbf{X}_-\phi_{1} =0,
 \label{ConstraintGravEq6}
 \\ &
  \tau\bm\partial_{\tau}\phi_{3}-\rho\bm\partial_{\rho}\phi_{3}
+ \frac{1}{2}\mathbf{X}_+ \phi_{4}- \frac{1}{2}\mathbf{X}_-\phi_{2}=0. 
\label{ConstraintGravEq7}
\end{align}
\end{subequations}

\begin{remark}
{\em The above evolution and constraint equations correspond,
  respectively, to equations (29a)-(29e) and (30a)-(30c) in \cite{GasVal20} with the
  replacement
\[
\eth \mapsto \mathbf{X}_+, \qquad \bar\eth\mapsto \mathbf{X}_-,
\]
where $\eth$ and $\bar{\eth}$ are the eth and ethbar operators of the
Newman-Penrose formalism ---see e.g. \cite{PenRin84}. Observe, however, that the equations in
\cite{GasVal20} are defined over the (conformal) spacetime manifold
whereas \eqref{EvolutionGravEq0}-\eqref{EvolutionGravEq4} and
\eqref{ConstraintGravEq5}-\eqref{ConstraintGravEq7} are defined on the
fibre space discussed in Section \ref{Section:FibreSpace}.
 }
\end{remark}

\subsection{Asymptotic expansions}
The fact that $\phi_n$ with $n = 0,1,2,3,4,$ have a well defined spin
weight allows to encode the angular dependence in terms of
spin-weighted spherical harmonics ---see \cite{Ste91,
  GasVal20,Val03a}. When working on the fibre space of Section
\ref{Section:FibreSpace} one makes use of the functions
$T_i{}^j{}_k:SU(2)\rightarrow \mathbb{C}$
introduced in \cite{Fri98a} rather than the usual functions ${}_s
Y_{lm}$. 

\medskip
In the following, consistent with the discussion in
\cite{Fri98a,Val03a,GasVal20} we consider solutions to the equations  \eqref{EvolutionGravEq0}-\eqref{EvolutionGravEq4} and
\eqref{ConstraintGravEq5}-\eqref{ConstraintGravEq7} of the form
\begin{equation}
 \phi_{n}=
 \sum_{p=|2-n|}^{\infty}\sum_{\ell=|2-n|}^{p}\sum_{k=0}^{2\ell}
 \frac{1}{p!}a_{n,p;\ell,m}(\tau) T_{2p}{}^k{}_{\ell-n}
 \rho^{p}, \label{ExpansionGravPhin}
\end{equation}
where $a_{n,p;\ell,m}:\mathbb{R}\rightarrow\mathbb{C}$ and
$n=0,\ldots,4$. The correspondence between the functions $T_i{}^j{}_k$ and the
harmonics ${}_s
Y_{lm}$ is given by
\[
{}_s Y_{nm} \mapsto (-\mbox{i})^{s+2n-m} \sqrt{\frac{2n+1}{4\pi}}
T_{2n}{}^{n-m}{}_{n-s}
\]
---see \cite{FriKan00} for more details. The expression \eqref{ExpansionGravPhin} is taken as an Ansatz for the
solution.  The convergence of solutions to the spin-2 equations of the
form given by \eqref{ExpansionGravPhin} has been analysed in
\cite{Fri03b}. Ansatz \eqref{ExpansionGravPhin} allows to reduce the
problem of constructing asymptotic expansions to the spin-2 equations
to the analysis of a set of ordinary differential equations
for $a_{n,p,\ell,m}(\tau)$. To see this, let 
\begin{equation}\label{NotationPhinCoef}
\phi_{n}^{(p)} \equiv\frac{\partial^{p}\phi_{n}}{\partial
  \rho^{p}}\Bigg|_{\rho=0},
\end{equation}
 with $n=0,1,2,3,4$. Taking the $p$-th derivative of
 equations \eqref{EvolutionGravEq0}-\eqref{ConstraintGravEq7} respect to
 $\rho$ and evaluating at the cylinder $\mathcal{I}$ gives the relations 
\begin{subequations}
\begin{align}
& (1+\tau)\bm\partial_{\tau}\phi_{0}^{(p)} +
  \mathbf{X}_+\phi_{1}^{(p)} +(p-2)\phi_{0}^{(p)}=0, \label{pthDerivativeGravEq0}\\ &
  \bm\partial_{\tau}\phi_{1}^{(p)} + \frac{1}{2} \mathbf{X}_-\phi_{0}^{(p)}+
  \frac{1}{2}\mathbf{X}_+\phi_{2}^{(p)} + \phi_{1}^{(p)}=0, \label{pthDerivativeGravEq1}\\ &
  \bm\partial_{\tau}\phi_{2}^{(p)}  +\frac{1}{2}\mathbf{X}_-\phi_{1}^{(p)}+\frac{1}{2}
\mathbf{X}_+ \phi_{3}^{(p)}=0,
 \label{pthDerivativeGravEq2}
  \\ &
  \bm\partial_{\tau}\phi_{3}^{(p)} +\frac{1}{2}\mathbf{X}_-\phi_{2}^{(p)}
+\frac{1}{2}\mathbf{X}_+\phi_{4}^{(p)}-\phi_{3}^{(p)}=0, \label{pthDerivativeGravEq3}\\ &
  (1-\tau)\bm\partial_{\tau}\phi_{4}^{(p)} + \mathbf{X}_- \phi_{3}^{(p)}  
+ (p- 2)\phi_{4}^{(p)}=0, \label{pthDerivativeGravEq4} 
\end{align}
\end{subequations}
and
\begin{subequations}
\begin{align}
&  \tau\bm\partial_{\tau}\phi_{1}^{(p)}  + \frac{1}{2} \mathbf{X}_+\phi_{2}^{(p)} - \frac{1}{2}\mathbf{X}_-\phi_{0}^{(p)}-p\phi_{1}^{(p)} =0, \label{pthDerivativeGravEq5}
\\ &
  \tau\bm\partial_{\tau}\phi_{2}^{(p)} 
 + \frac{1}{2}\mathbf{X}_+ \phi_{3}^{(p)} - 
\frac{1}{2}\mathbf{X}_-\phi_{1}^{(p)}-p\phi_{2}^{(p)} =0, \label{pthDerivativeGravEq6}
 \\ &
  \tau\bm\partial_{\tau}\phi_{3}^{(p)} 
+ \frac{1}{2}\mathbf{X}_+ \phi_{4}^{(p)}- \frac{1}{2}\mathbf{X}_-\phi_{2}^{(p)} 
-p\phi_{3}^{(p)}=0. 
\label{pthDerivativeGravEq7}
\end{align}
\end{subequations}
These equations, in turn, upon substitution of the expansion 
\eqref{ExpansionGravPhin}, imply a system of ordinary differential
equations for $a_{n,p;\ell,m}$ with
$p \geq 2$ and $2 \leq \ell \leq p$.
In fact, part of these equations constitutes an algebraic
system that allows to determine
$a_{1,p,\ell,m}$, $a_{2,p,\ell,m}$ and $a_{3,p,\ell,m},$
in terms of $a_{0,p,\ell,m}$ and $a_{4,p,\ell,m}$.
The problem is then reduced to solve the following system
of equations for  $a_{0,p,\ell,m}$ and $a_{4,p,\ell,m}$ 
\begin{subequations}
\begin{eqnarray}
& (1-\tau^2)\ddot{a}_{0}+(4+ 2(p-1)\tau)\dot{a}_{0} + (p+ \ell)(p-\ell
  + 1)a_{0}=0,
\label{SecondOrderGrav0} \\
& (1-\tau^2)\ddot{a}_{4}+(-4+ 2(p-1)\tau)\dot{a}_{4} + (p+
\ell)(p-\ell + 1)a_{4}=0.
\label{SecondOrderGrav4}
\end{eqnarray}
\end{subequations}
Further details of this calculation can be found in \cite{GasVal20, Val03a}.
Also, observe that if $a_{0}(\tau)$ solves \eqref{SecondOrderGrav0} 
then $a_{0}^{s}(\tau) \equiv a_{0}(-\tau)$ solves equation
\eqref{SecondOrderGrav4}.
Equations \eqref{SecondOrderGrav0} and \eqref{SecondOrderGrav4} are
examples of Jacobi differential equations. For $l \neq p$, the solution is
given by the so-called Jacobi polynomials as summarised in the
following:

\begin{lemma}\label{Spin2SolutionsExplicit}
  The solutions to the system \eqref{SecondOrderGrav0} and
  \eqref{SecondOrderGrav4} can be written as
\begin{align*}
  a_{0,p;l,m}(\tau) & = C_{p,\ell,m}Q^1_{p,\ell}(\tau) +
  (-1)^{\ell}D_{p,\ell,m}Q^3_{p,\ell}(\tau), \\ a_{4,p;l,m}(\tau)& =
  D_{p,\ell,m}Q^1_{p,\ell}(-\tau) +
  (-1)^{\ell}C_{p,\ell,m}Q^3_{p,\ell}(-\tau),
\end{align*}
with 
\[
C_{p,\ell,m} \equiv  X^{-1}_Aa_{0,p;l,m}(0) + X^{-1}_B
  a_{4,p;l,m}(0),  \qquad  D_{p,\ell,m} \equiv  X^{-1}_Ba_{0,p;l,m}(0) +
  X^{-1}_A a_{4,p;l,m}(0),
\]
where $Q_{p,\ell}^{n}(\tau)$ denotes some not everywhere vanishing Jacobi polynomials while
$X^{-1}_A$ and $X^{-1}_B$ are constant factors arising from evaluating
the Jacobi polynomials at $\tau=0$.
\end{lemma}

For further details see \cite{GasVal20}. The case $l=p$ is special and
the central point in the discussion of this section:

\begin{proposition}\label{Spin2LogarithmicTerms}
  For  $p\geq 2$, $\ell=p$, $-p \leq m \leq p$ the solution
  to equations  \eqref{SecondOrderGrav0} and
  \eqref{SecondOrderGrav4}  is given by
\begin{align}
a_{0,p;p,m}(\tau)= \left(\frac{1-\tau}{2}\right)^{p+2}
\left(\frac{1+\tau}{2}\right)^{p-2} \left( E_{p,m}+
E^{\ast}_{p,m}\int_{0}^{\tau}\frac{\mbox{ds}}{(1+s)^{p-1}(1-s)^{p+3}}
\right),\\ a_{4,p;p,m}(\tau)= \left(\frac{1+\tau}{2}\right)^{p+2}
\left(\frac{1-\tau}{2}\right)^{p-2} \left( I_{p,m}+
I^{\ast}_{p,m}\int_{0}^{\tau}\frac{\mbox{ds}}{(1-s)^{p-1}(1+s)^{p+3}}
\right).
\end{align}
where $E_{p,l,m}$, $E^{\ast}_{p,l,m}$ and $I_{p,l,m}$,
$I^{\ast}_{p,l,m}$ are integration constants.
\end{proposition}

\begin{remark}
{\em Observe that if $E^{\ast}_{p,l,m}$ and $I^{\ast}_{p,l,m}$ vanish then the
solutions $a_{n,p,\ell,m}$ are polynomial. To understand the
effect of non-vanishing  constants $E^{\ast}_{p,l,m}$ and
$I^{\ast}_{p,l,m}$, 
observe that using partial fractions, the integrals of Proposition
\ref{Spin2LogarithmicTerms} give
rise to logarithmic terms in the solution. More precisely, one has
that 
\begin{eqnarray*}
&&\int_0^\tau\frac{ds}{(1\pm s)^{p-1}(1\mp
  s)^{p+3}}=A_{*}\ln(1-\tau)+\frac{A_{p\pm 2}}{(1-\tau)^{p\pm
    2}}+\cdots+\frac{A_{1}}{(1-\tau)}+A_0
\nonumber \\
&&\phantom{\int_0^\tau\frac{ds}{(1+s)^{p-1}(1-s)^{p+3}}=}
+B_{*}\ln(1+\tau)+\frac{B_{p\mp 2}}{(1+\tau)^{p\mp 2}}+\cdots+\frac{B_{1}}{(1+\tau)},
\end{eqnarray*}
where $A_i$ and $B_i$ are constants.}
\end{remark}

The integration constants $E^{\ast}_{p,l,m}$ and $I^{\ast}_{p,l,m}$
associated to the logarithmic singularities in the
solutions for the $\ell=p$ modes can be related to initial data via the
linearisation of the Bach spinor $B_{ABCD}$ ---see \cite{Val03a}.
This gives rise to the following regularity condition ensuring the
vanishing of $E^{\ast}_{p,l,m}$ and $I^{\ast}_{p,l,m}$:

\begin{proposition}
  \label{prop:regularity_condition}
  The solution to the spin-2 equation on
  $(\mathcal{M},\bmeta)$ extends analytically to the critical sets $\mathcal{I}^\pm$ if and only
  if on the initial hypersurface
\[
\mathcal{S}_\star \equiv \big\{ p\in \mathcal{M} \; | \; \tau(p)=0 \big\},
\]
the \emph{regularity condition}
\begin{equation}
\label{regcond}
D_{(A_sB_s\cdots}D_{A_1B_1}B_{ABCD)}(i)=0, \quad \mbox{holds for} \quad s=0,1,\dots \;\; ,
\end{equation}
where $B_{ABCD}$ is the linearised Bach spinor, which can be
written in terms of the conformal factor and the spin-2 field as
\begin{equation}\label{BachSpinorDef}
B_{ABCD}=2D_{E(A}\Omega\phi_{BCD)}^{\phantom{BCD)}E}+\Omega
D_{E(A}\phi_{BCD)}^{\phantom{BCD)}E},
\end{equation}
where $D_{AB}$ is the spinorial counterpart of the Levi-Civita
connection of $(\bmh,\mathcal{S})$ where $\bmh$ is the
pull back of $\bmeta$ to $\mathcal{S}_\star$ and $\Omega\equiv \Theta|_{\mathcal{S}_\star}=\rho$.
\end{proposition}

\begin{remark}
\label{Remark:TaylorExp}
{\em It is important to stress that the regularity condition
  \eqref{regcond} does not imply, say, that $B_{ABCD}=0$ in a
  neighbourhood of $i$. This can be seen in the case that $B_{ABCD}$
  is analytic in a neighbourhood of infinity. In that case, the
  coefficients $D_{(A_sB_s\cdots}D_{A_1B_1}B_{ABCD)}(i)$ involve only
  a part of the derivatives at order $p$ ---the other parts involve a
  curl and a divergence. For example, if $p=1$ one has that the
  various components of $D_{A_1B_1} B_{ABCD}(i)$ are given by
\[
D_{(A_1B_1} B_{ABCD)}(i), \qquad D^{AB}B_{ABCD}(i), \qquad \mbox{and} \qquad D^Q{}_{(A}B_{BCD)Q}(i).
\]
A more detailed discussion of this issue and the relation between the
irreducible decomposition of spinors and Taylor expansions can be
found in \cite{Fri98a}, Section 3.3.
}
\end{remark}

\begin{remark}
{\em The regularity condition \eqref{regcond} is conformally
  invariant. This follows from the transformation of the linearised
  Bach spinor and the fact that it is a condition at a point ---see
  e.g. \cite{CFEBook}, Section 19.3 where the conformal properties of
  the analogue condition for the (non-linearised) Bach spinor are discussed.}
\end{remark}

Further discussion of the consequences of logarithmic singularities in
the asymptotic expansions of the spin-2 fields at the critical sets and
its connection to the (non) peeling properties of
$\phi_{ABCD}$ can be found in \cite{Val03a}. The non-linear version (for
the full conformal Einstein field equations) of the regularity
condition \eqref{regcond} was derived in \cite{Fri98a}.  The
polyhomogeneous peeling behaviour consistent with the existence of
logarithmic singularities in the rescaled Weyl spinor and the Weyl
spinor are given in
\cite{Fri98a,Val04d, Val04a, Val07a} and \cite{GasVal17a}
respectively.

\section{The space-spinor formalism}
\label{SpaceSpinorFormalism}

In this section we briefly recall the space-spinor formalism and its
naturally associated 1+3 split ---see
e.g. \cite{Som80,CFEBook}. Although similar in spirit, this formalism
is 
different in some key aspects from the more common 3+1 split. The most fundamental difference being that the
1+3 split is adapted to a preferred direction given by a vector field  $\tau^a$ instead of a
foliation.  In other words, the vector $\tau^a$ is, in general, not
hypersurface orthogonal and hence, the distribution 
generated by $\tau^a$ is not integrable.  The subsequent analysis
will further refinement of a $1+1+2$ spinorial split. This is analogous to the
$1+3$ split when an additional direction $\rho^a$ is singled out
---see \cite{CFEBook,Som80,Sza94a,Sza94b, KopVal21} for further
discussion on the space-spinor formalism, the $1+3$ split, the $1+1+2$
split and applications.  The basic elements of this formalism will be
discussed on a general manifold with metric $(\mathcal{M},\bmg)$ and
towards the end of this section, some results specific for Friedrich's
representation of the spatial infinity of the Minkowski spacetime
$(\mathcal{M},\bmeta)$ will be given.

\subsection{The general formalism}

Let $(\mathcal{M},\bmg)$ be a manifold equipped with a Lorentzian
metric and let $\{e_{\bma}{}^a\}$ denote an orthonormal frame. That
is, one has 
\[
g^{ab}= \eta^{\bma\bmb}e_{\bma}{}^ae_{\bmb}{}^b,
\]
where $\eta^{\bma\bmb}\equiv \text{diag}(1,-1,-1,-1)$.  An associated null
frame $\{l^a,n^a,m^a,\bar{m}^a\}$ can be constructed via
\begin{equation}\label{orthoframeTonullframe}
l^a = \frac{\sqrt{2}}{2}(\bme_{\bm0}{}^a +\bme_{\bm3}{}^a ), \qquad
n^a =\frac{\sqrt{2}}{2}(\bme_{\bm0}{}^a -\bme_{\bm3}{}^a ), \qquad m^a
= \frac{\sqrt{2}}{2}(\bme_{\bm1}{}^a -\mbox{i}\bme_{\bm2}{}^a ).
\end{equation}
The spinorial counterpart of this null frame is denoted by 
$\{ \bme_{\bmA\bmA'}{}^a\}$ and satisfies
\[
g^{ab}=\epsilon^{\bmA
  \bmB}\epsilon^{\bmA'\bmB'}e_{\bmA\bmA'}{}^a e_{\bmB\bmB'}{}^b.
\]
Let  $\tau^{AA'}$ represent the spinor counterpart of the timelike
vector $\sqrt{2} e_{\bm0}{}^a$. Let $\{\epsilon_{\bmA}{}^{A}\}$ be
a spinor dyad $\epsilon_{\bm0}{}^{A}= o^{A}, \epsilon_{\bm1}{}^{A}= \iota^{A}$
such that
\begin{equation}\label{eq_tau_spinor_dyad}
\tau^{A A'}=\epsilon_{\bm0}{}^{A}\epsilon_{\bm0'}{}^{A'} 
+ \epsilon_{\bm1}{}^{A}\epsilon_{\bm1'}{}^{A'}.
\end{equation}
The latter implies that the spinor $\tau^{AA'}$ satisfies
\[
\tau_{AA'}\tau^{BA'}= \epsilon_{A}{}^{B}.
\]
Consequently, the normalisation is fixed so that $\tau^{AA'}\tau_{AA'}= 2$.
This normalisation is consistent with the conventions of
\cite{Fri91}.

 The Hermitian spinor $\tau^{AA'}$ induces a
notion of Hermitian conjugation as follows: given a spinor $\mu_{AA'}$
its Hermitian conjugate $\mu^\dagger_{AA'}$ is defined as
\begin{equation}
\label{HermitianConjugation}
 \mu^{\dagger}_{CD} \equiv\tau_{C}{}^{A'} \tau_{D}{}^{A}
 \overline{\mu_{A A'}} = \tau_{C}{}^{A'} \tau_{D}{}^{A}
 \overline{\mu}_{A' A},
\end{equation}
 where the bar denotes complex conjugation. This definition is
 extended to higher valence spinors by requiring that $(\bmpi
 \bmmu)^\dagger=\bmpi^\dagger\bmmu^\dagger$.

\subsubsection{The space-spinor split}
 The Hermitian spinor $\tau^{AA'}$ 
induces a \emph{space-spinor split}
as follows: given a spinor $v_{AA'}$ its \emph{space-spinor decomposition}
is determined by 
\begin{equation}\label{standardspacespinorsplit}
v_{AA'}= \frac{1}{2}\tau_{AA'}v - \tau^{Q}{}_{A'}v_{(QA)},
\end{equation}
where $v \equiv v_{QQ'}\tau^{QQ'}$ and $v_{AB} \equiv
\tau_{B}{}^{A'}v_{AA'}$. Similar decompositions apply to higher valence spinors.  The
space-spinor split of a spinor $\bm\mu$ with $n$ unprimed indices and
$m$ primed indices $\mu_{A_1,..A_n,A',...,A'_{m}}$ can be succinctly
described as the process of transvecting it with
$\tau_{B_{1}}{}^{A'_{1}} \cdots \tau_{B_{m}}{}^{A'_{m}}$ to produce
its \emph{space-spinor counterpart} $\mu_{A_1,...A_n,B_1,..B_m}$, which
in turn is (irreducibly) decomposed in terms of totally symmetric
spinors of equal or lower valence ---see \cite{CFEBook} for a
comprehensive discussion.

\subsubsection{The space-spinor decomposition of the connection}
Any general connection $\breve{\bmnabla}$ ---not necessarily
the Levi-Civita for which the symbol $\bmnabla$ is reserved---
can be split as
\[ 
\breve{\nabla}_{A A'}= \frac{1}{2}\tau_{A
  A'}\mathcal{\breve{D}}-\tau_{A'}{}^{Q}\mathcal{\breve{D}}_{AQ},
\]
where
\[
\mathcal{\breve{D}}\equiv\tau^{AA'}\breve{\nabla}_{AA'} \qquad \text{and}
\qquad \mathcal{\breve{D}}_{AB}\equiv
\tau_{(B}{}^{A'}\breve{\nabla}_{A)A'},
\]
denote, respectively, the \emph{Fermi covariant derivative} in the direction
of $\tau^{AA'}$
and the \textit{Sen connection} of $\breve{\bm\nabla}$ relative to
$\tau^{AA'}$.  It is worth mentioning that even in the case of a
Levi-Civita connection $\bm\nabla$, the associated Sen connection
$\mathcal{\bmD}$ will have a non-vanishing torsion which is encoded in the derivatives of
$\tau^a$. In fact, as briefly described in the following, these derivatives play
an important role in the space-spinor decomposition of the components
of the connection.

\medskip
In the following, it is convenient to define 
\[
\mathcal{X}_{ABCD} \equiv
\frac{1}{\sqrt{2}}\tau_{D}{}^{Q'}\tau_{B}{}^{A'}\nabla_{AA'}\tau_{CC'}.
\]
This spinor, encoding the derivatives of $\tau^a$, can be
decomposed in terms of the reduced spinors
\[
\chi_{AB} \equiv
\frac{1}{\sqrt{2}}\tau_{B}{}^{A'}\mathcal{D}\tau_{AA'}, \qquad
\chi_{ABCD} \equiv
\frac{1}{\sqrt{2}}\tau_{D}{}^{C'}\mathcal{D}_{AB}\tau_{CC'}.
\]
The acceleration of $\tau^a$ is encoded in $\chi_{AB}$ while
$\chi_{ABCD}$ is the \emph{Weingarten spinor associated} to $\tau^a$.  If $\mathcal{X}^{Q}{}_{(AB)Q}=0$ then $\tau^a$ is
hypersurface orthogonal and the Levi-Civita connection induced on the
leaves of the associated foliation, to be denoted by
$D_{AB}$, is related to the Sen
derivative by
\begin{equation}\label{LeviCivitaToSen}
D_{AB}\mu_C = \mathcal{D}_{AB}\mu_C +
\frac{1}{\sqrt{2}}\chi_{ABC}{}^{Q}\mu_{Q}.
\end{equation}

\medskip
Let $\Gamma_{\bmA\bmA'\bmC \bmD}$ denote the (spin)
connection coefficients associated to the Levi-Civita connection
$\bm\nabla$ in the spin dyad $\{\epsilon_{\bmA}{}^{A}\}$. Defining
$\Gamma_{\bmA\bmB\bmC\bmD} \equiv
\tau_{\bmB}{}^{\bmA'}\Gamma_{\bmA\bmA'\bmC\bmD}$ and exploiting the
Hermitian conjugation operation, it can be shown that the spinor
$\mathcal{X}_{ABCD}$ encodes the \emph{real part} of the
connection. More precisely, one has that 
 \[
\mathcal{X}_{ABCD}=-\frac{1}{\sqrt{2}}(\Gamma_{\bmA\bmB\bmC\bmD}
+ \Gamma_{\bmA\bmB\bmC\bmD}^{\dagger}).
\]
The spinor $\xi_{ABCD}$ encoding the
\emph{imaginary part} of the connection is given by
\[
\xi_{\bmA\bmB\bmC\bmD}=\frac{1}{\sqrt{2}}(\Gamma_{\bmA\bmB\bmC\bmD} -
\Gamma_{\bmA\bmB\bmC\bmD}^{\dagger}).
\]

\subsubsection{The $1+1+2$ spinor decomposition}
Spinorial 1+1+2 splits naturally arise in settings where a
 spacelike vector $\rho^a$ is distinguished. Let $\rho^{AA'}$ denote
 the spinorial counterpart of $\rho^a$. In the following it is assumed
 that $\rho^{AA'}$ is orthogonal to $\tau^{AA'}$ ---that is,
\begin{equation}\label{rhotauOrtho}
  \tau_{AA'}\rho^{AA'}=0.
\end{equation}
 In analogy to equation
 \eqref{eq_tau_spinor_dyad} one can further specialise the dyad
 $\{\epsilon_{\bmA}{}^A\}$ so that 
\begin{equation}\label{eq_rho_spinor_dyad}
\rho^{A A'}=\epsilon_{\bm0}{}^{A}\epsilon_{\bm0'}{}^{A'} -
\epsilon_{\bm1}{}^{A}\epsilon_{\bm1'}{}^{A'}.
\end{equation}
This choice of dyad is consistent with equation
\eqref{eq_tau_spinor_dyad}, but reduces the freedom available in the
dyad to $U(1)$ transformations. As a consequence of equation \eqref{eq_rho_spinor_dyad} , the spinor $\rho^{AA'}$ satisfies
\[
\rho_{AA'}\rho^{BA'} =- \delta_{A}{}^{B}.
\]
The latter implies $\rho_{AA'}\rho^{AA'}=-2$ consistent with
the assumption that $\rho^a$ is a spacelike vector. Consistent with
the previous discussion we identify $\rho^a$ with $\sqrt{2}\bme_{\bm3}{}^{a}$.
Additionally, the orthogonality condition \eqref{rhotauOrtho},
implies that $\rho_{AB}=\rho_{(AB)}$ where
\[
\rho_{AB} \equiv \tau_{B}{}^{A'}\rho_{AA'}.
\]
Further discussion on the spinorial $1+1+2$ split can be found in
\cite{Sza94a,Sza94b, KopVal21}.

\medskip
Particularising the previous general discussion to Friedrich's
representation of the spatial infinity of the Minkowski spacetime and
using the lift to the fibre space of the null frame \eqref{Fframe00}-\eqref{Fframe0110} and exploiting \eqref{orthoframeTonullframe} one
can directly read
\begin{equation}\label{taurhoToPartials}
  \bm \tau= \sqrt{2}\bm\partial_\tau, \qquad
  \bm \rho=\sqrt{2}( \rho
  \bm\partial_\rho-\tau \bm\partial_\tau).
\end{equation} 
Using the latter expressions and the frame a
calculation carried out using the suite {\tt xAct } for tensor and
spinorial manipulations in {\tt Mathematica} ---see \cite{GarMar12}---
gives that 
\begin{align}\label{ConnectionForMinkCyl}
  \chi_{ABCD}=0, \qquad \chi_{AB}=-\frac{2}{\sqrt{2}}\rho_{AB}.
\end{align}
The expressions for the 
spin connection coefficients $\Gamma_{\bmA\bmA'}{}^{\bmB}{}_{\bmC}$ in
the extended bundle space framework were given in Section \ref{Section:FibreSpace}.
\section{The staticity condition}
\label{Section:Staticity}

In this section we formulate a suitable staticity condition for spin-2
fields in the Minkowski spacetime. We then proceed to express this condition
in terms of the corresponding rescaled fields
in the conformal representation of the Minkowski spacetime
discussed in Section \ref{CylinderAtspatialInfinity}.
The latter will allow in turn to relate the staticity condition
with the regularity condition discussed in Section \ref{RegularityCondition}.

\subsection{Staticity conditions in the physical spacetime}
\label{staticity_propagation_physical}
 Let $\tilde{\phi}_{abcd}$ be a spin-2 field on Minkowski spacetime
 $(\mathbb{R}^4,\tilde{\bm\eta})$.  In other words, let
 $\tilde{\phi}_{abcd}$ be a tensor satisfying
\begin{equation}\label{TensorialSpin2FieldPhysical}
\tilde{\nabla}^a\tilde{\phi}{}_{abcd}=0.
\end{equation}
where $\tilde{\phi}_{abcd}$ possesses the same symmetries of the Weyl
tensor. Its spinorial counterpart can be decomposed as
\begin{equation}\label{IrrDecotildephi}
  \tilde{\phi}_{AA'BB'CC'DD'}=\tilde{\phi}_{ABCD}\tilde{\epsilon}_{A'B'}
  \tilde{\epsilon}_{C'D'}
+\bar{\tilde{\phi}}_{A'B'C'D'}
\tilde{\epsilon}_{AB}\tilde{\epsilon}_{CD},
\end{equation}
where $\tilde{\phi}_{ABCD}=\tilde{\phi}_{(ABCD)}$. The spinor field $\tilde{\phi}_{ABCD}$ satisfies
\begin{equation}\label{SpinorialSpin2FieldPhysical}
\tilde{\nabla}_{A'}{}^{A}\tilde{\phi}_{ABCD}=0.
\end{equation}

\subsubsection{Definition}
Consistent with the notation of Section
\ref{CylinderAtspatialInfinity}, let
$\tilde{x}^{\mu}=(\tilde{t},\tilde{x}^{i})$ denote Cartesian
coordinates. In the following we make use of the following \emph{ad hoc} notion of staticity for
the spin-2 field:

\begin{definition}
\label{Definition:Staticity}
The spin-2 field $\tilde{\phi}{}_{abcd}$ on the Minkowski spacetime is
static if and only if
\[
\mathcal{L}_{\tilde{\xi}}\tilde{\phi}_{abcd}=0, \qquad \mbox{with} \qquad
\tilde{\bm\xi}=\bm\partial_{\tilde{t}}.
\]
\end{definition}

In view of the latter definition it is convenient to introduce the \emph{zero-quantity}
\[
\tilde{Z}_{abcd}\equiv
\mathcal{L}_{\tilde{\xi}}\tilde{\phi}_{abcd}, \qquad  \mbox{with} \qquad
\tilde{\bm\xi}=\bm\partial_{\tilde{t}}.
\]
Expressing the Lie derivative in terms of covariant derivatives
and using the fact that the Killing vector $\bm\partial_{\tilde{t}}$
in the Minkowski spacetime is covariantly constant so that
$\tilde{\nabla}_{a}\tilde{\xi}^{e}=0$, one concludes that 
\begin{equation}
\label{DefZ}
\tilde{Z}_{abcd}=\tilde{\xi}^{e}\tilde{\nabla}_{e}\tilde{\phi}_{abcd}.
\end{equation}

\subsubsection{Propagation equations}
The next step in our analysis is to formulate the \emph{spacetime
  staticity notion} encoded in Definition \ref{Definition:Staticity}
in terms of conditions on the initial data for the spin-2 equation. 

\medskip 
Using equation \eqref{TensorialSpin2FieldPhysical} and
exploiting the fact that
\[
[\mathcal{L}_{\tilde{\xi}},\tilde{\nabla}]\bmT=0
\]
 for any Killing
vector $\tilde{\bm\xi}$ and any tensor field $\bmT$ ---see
\cite{Wei90a}--- one concludes that
\begin{equation}
\label{EquationForZPhysical}
\tilde{\nabla}^a\tilde{Z}{}_{abcd}=0.
\end{equation}
It can be readily verified that
the spinorial version of
equation \eqref{EquationForZPhysical} is 
\begin{equation}
\label{Spin-2EqZPhysical}
\tilde{\nabla}_{A'}{}^{A}\tilde{Z}_{ABCD}=0,
\end{equation}
where $\tilde{Z}_{AA'BB'CC'DD'}$ is the spinorial counterpart of
$\tilde{Z}_{abcd}$ and is related to $Z_{ABCD}$ via
\begin{equation}
\label{ZspinorIrrDecomp}
  \tilde{Z}_{AA'BB'CC'DD'}=\tilde{Z}_{ABCD}\tilde{\epsilon}_{A'B'}
  \tilde{\epsilon}_{C'D'}
+\bar{\tilde{Z}}_{A'B'C'D'}
\tilde{\epsilon}_{AB}\tilde{\epsilon}_{CD}.
\end{equation}

In order to extract the content of equation \eqref{Spin-2EqZPhysical}
we make use of the space-spinor formalism. Setting $\tilde{\tau}^{AA'} = \sqrt{2}\tilde{\xi}^{AA'}$
where $\tilde{\xi}^{AA'}$ denotes the spinorial counterpart of
$\tilde{\xi}^a$, a direct calculation using the space-spinor formalism described
in Section \ref{SpaceSpinorFormalism} shows that equation
\eqref{Spin-2EqZPhysical} can be split as
\begin{subequations}
\begin{eqnarray}
  &&\tilde{\mathcal{D}}\tilde{Z}_{ABCD}-2
  \tilde{\mathcal{D}}^{Q}{}_{(A}\tilde{Z}_{BCD)Q}=0,
\label{EvoPhysicalZ}\\
&&\tilde{\mathcal{D}}^{AB}\tilde{Z}_{ABCD}=0.\label{ConstPhysicalZ}
\end{eqnarray}
\end{subequations}
with $\tilde{\mathcal{D}}$ and $\tilde{\mathcal{D}}_{AB}$ as defined
in Section \ref{SpaceSpinorFormalism}.  The evolution equations
\eqref{EvoPhysicalZ} imply a symmetric hyperbolic system for the
components of $\tilde{Z}_{ABCD}$. This observation follows by formal
analogy to the analysis of the hyperbolic reductions of the 
spin-2 equation ---see \cite{CFEBook} for further discussion on the
hyperbolicity analysis of this type of spinorial equations.
Consequently, exploiting the standard uniqueness result for
symmetric hyperbolic systems, one has that
\begin{equation}
\label{KeyImplicationZPhysical}
\tilde{Z}_{ABCD}|_{\tilde{\mathcal{S}}}=0 \quad \Longleftrightarrow \quad
\tilde{Z}_{ABCD}=0 \qquad \text{in} \qquad D^+(\tilde{\mathcal{S}}), 
\end{equation}
with $\tilde{\mathcal{S}}$, say,  a Cauchy hypersurface of the
Minkowski spacetime. 

\medskip
In summary, it follows from the previous discussion that the requirement of
staticity for the spin-2 field on the Minkowski spacetime (as given by
Definition \ref{Definition:Staticity}) is equivalent to the initial
data condition  
\begin{equation}\label{staticity_cond}
\tilde{Z}_{ABCD}|_{\tilde{\mathcal{S}}}=0.
\end{equation}
In the following we refer to \eqref{staticity_cond} as to the
\emph{staticity condition}.  Condition \eqref{staticity_cond} will impose some
restrictions on the initial data for the physical spin-2 field
$\tilde{\phi}_{ABCD}$. In the following we analyse the relation between
this condition and the regularity condition of Proposition
\ref{prop:regularity_condition}. In order to do so, the two conditions
needs to be written in the same framework. 

\subsection{Staticity condition in the unphysical spacetime}
\label{StaticityUnphysicalSpacetime}

In this subsection we recall general conformation formulate. These expressions are then
particularised to the case of the Minkowski spacetime
$(\mathbb{R}^4,\tilde{\bm\eta})$ and the conformal extension
$(\mathcal{M},\bmeta)$ discussed in Section
\ref{CylinderAtspatialInfinity}.  Finally, the staticity condition
\eqref{staticity_cond} is recast in terms of the (unphysical) spin-2
field $\phi_{ABCD}$ propagating in $(\mathcal{M},\bmeta)$.

\subsubsection{Conformal transformation formulae}

We start by recalling that:

\begin{lemma}
  \label{conf_invariance_spin2}
Let $(\mathcal{M},\bmg)$ and
$(\tilde{\mathcal{M}},\tilde{\bmg})$ two Lorentzian manifolds equipped
metrics related through a conformal transformation
\begin{equation}\label{conf_resc}
  \bmg = \Xi^2 \tilde{\bmg}.
\end{equation}
Let $\tilde{\phi}_{abcd}$ be a tensor with the same symmetries of the
Weyl tensor then,
\begin{equation}
\tilde{\nabla}^a\tilde{\phi}{}_{abcd}=0 \implies \nabla^a\phi{}_{abcd}=0,
\end{equation}
where $\tilde{\phi}_{abcd}=\Xi^{-1}\phi_{abcd}$.  In terms of their
reduced spinorial counterparts $\tilde{\phi}_{ABCD}$ and $\phi_{ABCD}$
one similarly has that
\begin{equation}\label{spin2Trans}
\tilde{\nabla}_{A'}{}^{A}\tilde{\phi}_{ABCD}=0 \implies
\nabla_{A'}{}^{A}\phi_{ABCD}=0,
\end{equation}
where $\phi_{ABCD}=\Xi^{-1}\tilde{\phi}_{ABCD}$.
\end{lemma}

This classical result follows from a direct calculation using the
conformal transformation formulae relating the connections
$\bm\nabla$ and $\tilde{\bm\nabla}$---see \cite{Ste91, PenRin86}. We
also note the following:

\begin{lemma}
If $(\tilde{\mathcal{M}},\tilde{\bmg})$ admits a Killing
vector $\tilde{\bm\xi}$, namely,
$\tilde{\nabla}_{(a}\tilde{\xi}_{b)}=0,$ then $\tilde{\bm\xi}$ gives
rise to a conformal Killing vector $\bm\xi$ on
$(\mathcal{M},\bmg)$. That is, one has 
\begin{equation}
\label{CKVeq}
  \nabla_{(a}\xi_{b)} - \frac{1}{4}g_{ab}\nabla_c\xi^c=0,
  \qquad \text{where} \qquad
  \nabla_a\xi^a= 4 \Xi \xi^a\nabla_a \Xi,
\end{equation}
with
\begin{equation}
\label{KVtoCKV_general}
\xi^a = \tilde{\xi}^a, \qquad \xi_a = \Xi^{-2}\tilde{\xi}_a.
\end{equation}
\end{lemma}
This result is straightforwardly obtained by direct substitution ---see
also \cite{Pae14}.

\begin{remark}
  \emph{ The conformal factor appears in the second expression in
    equation \eqref{KVtoCKV_general} because the indices of tensors in
    $(\tilde{\mathcal{M}},\tilde{\bmg})$ are, by definition, raised
    and lowered using $\tilde{\bmg}$ while for tensors in
    $(\mathcal{M},\bmg)$, the metric $\bmg$ is used.  For the case of
    the spin-2 equation it is enough to identify explicitly the
    conformal Killing vector on $(\mathcal{M},\bmeta)$ associated to
    the timelike Killing vector in $(\mathbb{R}^4,\tilde{\bm\eta})$.
    However for more general spacetimes one would need to make
    use of the conformal Killing initial data equations of
    \cite{Pae14}.  }
\end{remark}

\subsubsection{Propagation equations in the unphysical spacetime}
Most of the subsequent discussion can be made general without having
to make use of the particular features of the conformal extension
$(\mathcal{M}, \bmeta)$ of the Minkowski spacetime.

\medskip
As before,  let $\tilde{Z}_{abcd} \equiv \mathcal{L}_\xi
\tilde{\phi}_{abcd}$. Using that  $\tilde{\phi}_{abcd}=\Xi^{-1}\phi_{abcd}$ one has that
\[
\tilde{Z}_{abcd} = \phi_{abcd}\mathcal{L}_\xi \Xi + \Xi \mathcal{L}_{\xi}\phi_{abcd}.
\]
A direct spinorial translation of the latter equation gives
\begin{equation}
\label{ZtoLiePhi_gen}
\tilde{Z}_{AA'BB'CC'DD'} = \phi_{AA'BB'CC'DD'}\mathcal{L}_\xi \Xi + \Xi \mathcal{L}_{\xi}\phi_{AA'BB'CC'DD'}.
\end{equation}

\begin{remark}
  \emph{It is worth making a word of caution regarding the notion of
    Lie derivative for spinorial fields. Give a spinor $\mu^A$, its Lie derivative
$\mathcal{L}_{\xi}\mu^A$ is only well defined when $\xi^a$ is a
conformal Killing vector ---see \cite{PenRin86}.  This is not a
problem in the current set up as $\xi^a$ is assumed from the outset
to be a conformal Killing vector. Moreover, the conformal Killing vector
equation can be expressed in spinorial terms as
\begin{equation}
\mathcal{L}_\xi \epsilon^{AB} = \lambda \epsilon^{AB},
\end{equation}
where
\begin{equation}\label{real_part_div}
\nabla_{AA'}\xi^{AA'} =-2 (\lambda + \bar{\lambda}).
\end{equation}
With this notation at hand the Lie-derivative along $\xi^a$ of a
valence-1 spinor $\mu^A$ is given by
\begin{equation}\label{LieDermu}
\mathcal{L}_\xi\mu^A = \xi^{QQ'}\nabla_{QQ'}\mu^A -{h}_{Q}{}^{A}\mu^Q,
\end{equation}
where
\begin{equation}
{h}_{A}{}^{B}= \frac{1}{2}(\epsilon_{A}{}^{B}\bar{\lambda} +
\nabla_{AQ'}\xi^{BQ'}),
\end{equation}
---see \cite{PenRin86} for further details.  As discussed in
\cite{PenRin86} the divergence of the conformal Killing vector only
fixes the real part of $\lambda$ and a geometrically natural choice  is to set $\lambda = \bar{\lambda}$
---consistent with a conformal rescaling using a real function $\Xi$
instead of a complex one.}
  \end{remark}

Applying $\mathcal{L}_\xi$ to the irreducible decomposition
for the spinorial counterpart of $\phi_{abcd}$ ---the analogue of
equation \eqref{IrrDecotildephi}--- gives
\begin{equation}
\label{Liephi}
  \mathcal{L}_\xi\phi_{AA'BB'CC'DD'}=\epsilon_{A'B'}\epsilon_{C'D'}
  \mathcal{L}_\xi\phi_{ABCD} -
  2\bar{\lambda}\phi_{ABCD}\epsilon_{A'B'}\epsilon_{C'D'} + c.c.,
\end{equation}
where $c.c.$ denotes the complex conjugate of the displayed
expression. Now, defining
\[
Z_{ABCD} \equiv \Xi^{-2}\tilde{Z}_{ABCD},
\]
and observing that 
$\epsilon_{AB}=\Xi\tilde{\epsilon}_{AB}$, consistent with the
conformal transformation \eqref{conf_resc}, gives
\begin{equation}
\label{ZIrrDecomps}
 \tilde{Z}_{AA'BB'CC'DD'}
 = Z_{ABCD}\epsilon_{A'B'}\epsilon_{C'D'} +\bar{Z}_{A'B'C'D'}
 \epsilon_{AB}\epsilon_{CD}.
\end{equation}
Substituting equations \eqref{Liephi} and \eqref{ZIrrDecomps} into
equation \eqref{ZtoLiePhi_gen} and symmetrising gives
\begin{equation}
\label{ZrescaledToLiephi}
Z_{ABCD} = \Xi\mathcal{L}_{\xi}\phi_{ABCD} - 2\Xi
\bar{\lambda}\phi_{ABCD} + \phi_{ABCD}\mathcal{L}_\xi \Xi.
\end{equation}
Unwrapping the expression for the Lie derivative
$\mathcal{L}_{\xi}\phi_{ABCD}$, using the generalisation of equation
\eqref{LieDermu} for a totally symmetric valance-4 spinor (see
\cite{PenRin86}) leads to 
\begin{equation}
\label{LiephiExpanded}
  \mathcal{L}_{\xi}\phi_{ABCD} = \xi^{QQ'}\nabla_{QQ'}\phi_{ABCD} + 2
  \phi_{Q(ABC}\nabla_{D)Q'}\xi^{QQ'} + 2 \bar{\lambda}\phi_{ABCD}.
\end{equation}
Accordingly, equation \eqref{ZrescaledToLiephi} can be rewritten in a more
explicit form as
\begin{equation}
\label{ZrescaledNoLie}
Z_{ABCD} = \Xi\xi^{QQ'}\nabla_{QQ'}\phi_{ABCD} +
\phi_{ABCD}\mathcal{L}_\xi \Xi + 2 \phi_{Q(ABC}\nabla_{D)Q'}\xi^{QQ'}.
\end{equation}
Notice that the term $\bar{\lambda}$ present in equation
\eqref{LiephiExpanded} is absent once one expands the Lie derivative
of $\phi_{ABCD}$.

\subsubsection{Space-spinor decomposition of the staticity condition}
In view of the eventual evaluation of $Z_{ABCD}$ on the
initial hypersurface $\mathcal{S}$, a 1+3 split of equation
\eqref{ZrescaledNoLie} is in order. Let
$\tau^{AA'}$ correspond to the vector $\sqrt{2}\bm\partial_\tau$ with
$\bmg(\bm\partial_\tau,\bm\partial_\tau)=1$.  To employ the
space-spinor split induced by $\tau^{AA'}$, define
\begin{equation}
\label{xinormaltangentialdef}
\xi \equiv \frac{1}{\sqrt{2}}\xi^{AA'}\tau_{AA'}, \qquad \xi_{AB}
\equiv \xi_{A'(A}\tau_{B)}{}^{A'},
\end{equation}
so that $\xi_{AA'}$ can be decomposed as
\begin{equation}
\label{ckvspacepinor_decomp}
\xi_{AA'} = \frac{1}{\sqrt{2}}\xi\tau_{AA'} + \xi_{A}{}^{B}\tau_{BA'}.
\end{equation}

\medskip
The decomposition \eqref{ckvspacepinor_decomp} and the space
spinor formalism can be employed to rewrite each of the terms in the
right hand side of equation \eqref{ZrescaledNoLie} as follows: for
the first term a short calculation gives
\begin{subequations}\label{terms_Z}
\begin{equation}\label{first_term_Z}
\xi^{FA'} \nabla_{FA'}\phi_{ABCD} = \tfrac{1}{\sqrt{2}}\xi \mathcal{D}
\phi_{ABCD} + \xi^{FG} \mathcal{D}_{GF}\phi_{ABCD},
\end{equation}
and for the second term
\begin{equation}
\label{second_term_Z}
  \mathcal{L}_\xi \Xi = \frac{1}{\sqrt{2}}\xi \mathcal{D}\Xi +
  \xi^{AB} \mathcal{D}_{AB}\Xi .
\end{equation}
For the last term in the righthand side of equation
\eqref{ZrescaledNoLie} a longer calculation shows that 
\begin{align}\label{third_term_Z}
 4\phi_{BCDF}{}^{F}\nabla_{AA'}\xi_{F}{}^{A'} & = \chi_{A}{}^{F} (2
 \xi_{F}{}^{G} \phi_{BCDG} - \sqrt{2} \phi_{BCDF} \xi) - 2 \phi_{BCDF}
 \mathcal{D}\xi_{A}{}^{F} + \sqrt{2} \phi_{ABCD} \mathcal{D}\xi
 \nonumber \\ & \hspace{-5mm} - 4 \phi_{BCDG}
 (\xi\chi_{A}{}^{F}{}_{F}{}^{G} + \sqrt{2} \xi^{FG}
 \chi_{A}{}^{H}{}_{FH} - \mathcal{D}_{AF}\xi^{FG}) + 2 \sqrt{2}
 \phi_{BCDF} \mathcal{D}_{A}{}^{F}\xi.
\end{align}
\end{subequations}
where $\chi_{ABCD}$ and $\chi_{AB}$ denote encode the tangential and
normal derivatives of $\tau^a$ as defined in Section
\ref{SpaceSpinorFormalism}. 

\begin{remark}
{\em Observe that expressions
\eqref{first_term_Z}-\eqref{third_term_Z} contain both normal and
tangential derivatives of $\xi$, $\xi_{AB}$ and $\phi_{ABCD}$.  The
normal derivatives can be replaced by tangential derivatives
exploiting the conformal Killing vector and the spin-2 equations
satisfied by $\xi_{AA'}$ and $\phi_{ABCD}$, respectively. More
precisely, one has that:
\begin{itemize}
\item[(a)] The direct spinorial translation of the conformal Killing
vector equation \eqref{CKVeq} reads
  \begin{equation}
\label{CKVeq_spinorial}
\nabla_{AA'}\xi_{BB'} + \nabla_{BB'}\xi_{AA'} -
\frac{1}{2}\epsilon_{AB}\epsilon_{A'B'}\nabla_{CC'}\xi^{CC'}=0.
  \end{equation}
  Substituting the decomposition \eqref{ckvspacepinor_decomp} into
  equation \eqref{CKVeq_spinorial}, using the space-spinor split and
  solving for the normal derivatives of $\xi$ and $\xi_{AB}$ one obtains
\begin{subequations}
  \begin{eqnarray}
    && \mathcal{D}\xi = \Xi \xi  \mathcal{D}\Xi +
    \frac{1}{\sqrt{2}}\chi^{AB}\xi_{AB} +
    \frac{2}{\sqrt{2}}\xi^{AB}\mathcal{D}_{AB}\Xi, \label{NormalDerivativesOfFields1}\\
&&
    \mathcal{D}\xi_{AP} = \frac{1}{2} \chi_{P}{}^{B} \xi_{AB} +
    \frac{1}{2} \chi_{A}{}^{B} \xi_{PB} -
    \frac{1}{\sqrt{2}}\chi_{AP} \xi + \sqrt{2} \xi^{BC} \chi_{APBC} -
    \sqrt{2} \mathcal{D}_{AP}\xi. \label{NormalDerivativesOfFields2}
\end{eqnarray}
\end{subequations}

\item[(b)]  Similarly, from the spin-2 equation \eqref{spin2Trans}, using the
  space-spinor formalism and solving for the normal derivatives of
  $\phi_{ABCD}$ one finds that 
  \[
\mathcal{P} \phi_{ABCD} = 2\mathcal{D}^{Q}{}_{(A}\phi_{BCD)Q}.
  \]

\end{itemize}
}
\end{remark}

 Using equations \eqref{NormalDerivativesOfFields1}-\eqref{NormalDerivativesOfFields2} to replace the
  normal derivatives in \eqref{terms_Z} and substituting this result
  into equation \eqref{ZrescaledNoLie} one obtains an expression which
  only contains tangential (Sen) derivatives.  This long
  expression is not very insightful. 

\subsubsection{Specific expressions for the Minkowski spacetime}
At this point it is
  convenient to particularise the discussion for the case of  the
  conformal extension $(\mathcal{M},\bmeta)$ of the Minkowski
  spacetime.

\medskip
The most important simplification that occurs in the present case, is
that the conformal Killing vector $\bmxi$ associated to
$\tilde{\bm\xi} = \bm\partial_{\tilde{t}}$ is, in fact, a Killing
vector of the unphysical spacetime
$(\mathcal{M},\bmeta)$. To see this, notice that using
the coordinate transformations \eqref{UnphysicalToPhysicaltrho} and
$t=\tau\rho$, the conformal factor $\Theta$ given in equation
\eqref{ThetaInUnphysicalCoords} when expressed in the physical coordinates
$(\tilde{t},\tilde{\rho})$ simply reads
\begin{equation}
\label{CFthetaToPhysCoords}
\Theta = -\frac{1}{\tilde{\rho}}.
\end{equation}
Observing that $\bmxi=\tilde{\bmxi}=\bm\partial_{\tilde{t}}$,
using equation \eqref{CFthetaToPhysCoords} and the second expression
in equation \eqref{CKVeq} one concludes that $\bm\xi =
\bm\partial_{\tilde{t}}$ is a Killing vector in
$(\mathcal{M},\bmeta)$. Moreover, a direct application
of the chain rule using equations \eqref{UnphysicalToPhysicaltrho} and
$t=\tau\rho$ gives 
\begin{equation}
\label{CKV_cylinder_mink}
\bm\xi = \bm\partial_{\tilde{t}}= \Theta \bm\partial_\tau +
2\rho^2\tau \bm\partial_\rho.
\end{equation}
Using the equation
\eqref{CKV_cylinder_mink}
one can identify the Killing shift and lapse $\xi$ and $\xi_{AB}$.  A
calculation observing 
\eqref{taurhoToPartials} gives
\begin{align}
\label{xiDecompMinkCyl}
\xi = 2 \rho \tau^2 + \Theta, \qquad \xi^{AB}= \sqrt{2}
\rho\tau\rho^{AB}.
\end{align}

\subsubsection{The final condition}
Observing that $\mathcal{S}\setminus i^0$ and $\tilde{\mathcal{S}}$
are diffeomorphic and using that $Z_{ABCD}=\Theta^{2}\tilde{Z}_{ABCD}$ the
staticity condition \eqref{staticity_cond} can be expressed as
\[
\tilde{Z}_{ABCD}|_{\tilde{\mathcal{S}}}=\Theta^{2}
Z_{ABCD}|_{\mathcal{S}\setminus i^0} = 0.
\]
Thus, by continuity it follows, in fact that 
\[
Z_{ABCD}|_{\mathcal{S}}=0.
\]
Using this formulation of the staticity condition and 
substituting equations \eqref{ConnectionForMinkCyl},
\eqref{xiDecompMinkCyl} along with
\eqref{NormalDerivativesOfFields1}-\eqref{NormalDerivativesOfFields2}
and \eqref{terms_Z} into equation \eqref{ZrescaledNoLie} one finds,
after a long computation that the staticity condition can be rewritten
in the very compact form
\begin{equation}
\label{staticitycondSen}
Z_{ABCD}|_{\mathcal{S}}=
\sqrt{2}\mathcal{D}_{(A}{}^{Q}\Big(\rho^2\phi_{BCD)Q}\Big)\Big|_{\mathcal{S}}
= 0.
\end{equation}

Although in the present case $\tau^a$ is not hypersurface orthogonal,
one nevertheless has that  \eqref{LeviCivitaToSen}
holds on $\mathcal{S}$. Hence, considering the pull back of equations
\eqref{ConnectionForMinkCyl} and \eqref{LeviCivitaToSen} to
$\mathcal{S}$ one has
\begin{equation}
\label{staticityLevi}
Z_{ABCD}|_{\mathcal{S}}=
\sqrt{2}D_{(A}{}^{Q}\Big(\rho^2\phi_{BCD)Q}\Big)\Big|_{\mathcal{S}}.
\end{equation}
Now, recalling that the linearisation of the Bach tensor is given
in terms of $\phi_{ABCD}$ and $\Omega \equiv \Theta|_{\mathcal{S}}$ by
\begin{equation}\label{Bachspinor}
  B_{ABCD}=D_{(A}{}^{Q}\Big(\Omega^2\phi_{BCD)Q}\Big).
\end{equation}
 Hence, using \eqref{staticityLevi} and
\eqref{Bachspinor} and the above observations one concludes that
\begin{equation}
\label{ZequalsB}
{Z}_{ABCD} =\sqrt{2} B_{ABCD}.
\end{equation}

\begin{remark}
{\em
It can be verified that the linearised Bach spinor $B_{ABCD}$
transforms homogeneously under conformal transformations. More
precisely, under the rescaling $\bmh\mapsto \phi^2 \bmh$ one has that 
\[
B'_{ABCD} \mapsto \phi^{-1}B_{ABCD}.
\]
It follows that the condition $B_{ABCD}=0$ on $\mathcal{S}$ is
conformally invariant under rescalings of the initial metric.}
\end{remark}

The discussion of this section can be summarised in the following:

\begin{theorem}
\label{Theorem:Main}
A necessary and sufficient condition for a spin-2 field $\phi_{ABCD}$
over $(\mathcal{M},\bmeta)$ to be static in the sense of Definition
\ref{Definition:Staticity} is that it satisfies the conformally
invariant condition
\[
B_{ABCD}=0 \quad \mbox{on} \quad \mathcal{S}. 
\]
\end{theorem}

It follows from the above result that any static spin-2 field trivially
satisfies the regularity condition of \eqref{regcond}. Accordingly,
one has the following:

\begin{corollary}
  Static initial data for the spin-2 field gives rise to a
  solution $\phi_{ABCD}$ that extends analytically to the the critical
  sets $\mathcal{I}^{\pm}$. In particular, the solution is smooth at
  $\mathscr{I}^\pm$. 
\end{corollary}

Moreover, consistent with Remark \ref{Remark:TaylorExp}, we have the
following main conclusion of the present analysis:

\begin{corollary}
  Initial data satisfying the regularity condition \eqref{regcond}
  does not correspond, in general, to \emph{static initial data} for the
  spin-2 field in a neighbourhood of $i$.
\end{corollary}

\section{Conclusions}
The objective of this article has been to analyse the relation between
a regularity condition on initial data for the spin-2 field which
ensures that the associated solutions extend smoothly (and in fact, analytically) through the critical
sets $\mathscr{I}^+$ where null infinity meets spatial infinity. Making use of the
estimates for the solutions to the spin-2 equation developed in
\cite{Fri03b} it follows that regularity at the critical sets implies
smoothness at null infinity, $\mathscr{I}^\pm$. In contrast to the case of the Einstein
field equations (the ultimate motivation of our analysis) where
smoothness at null infinity is closely related to the staticity of the
initial data (see e.g. \cite{Fri13,Fri18,Val10,Val11}), for the linear
spin-2 field there exists, a priori, large classes of data satisfying the
regularity condition on the linearised Bach spinor which are not
static. The underlying reason for this difference can be understood in
terms of the more stringent set of (nonlinear) constraints that the
initial data sets
for the Einstein field equations have to satisfy. Observe also, that
while the staticity condition introduced in Definition
\ref{Definition:Staticity} leads to a linear condition on the initial
data, the conformal static equations (see e.g. \cite{CFEBook}, Chapter
19) are a nonlinear system. 

The results of our analysis thus, serve as a cautionary note towards
inferring asymptotic properties of solutions to the Einstein field
equations from linearised models. One cannot expect to be able to
capture the whole richness of the Einstein field equations through the
analysis of linearised field equations.

Finally, we observe that a similar analysis can, in principle, be
carried out for other massless field equations (e.g. the Dirac and
Maxwell fields). Of more interest for the Einstein field equations is
the analysis of nonlinear systems like the Maxwell-Dirac or
Maxwell-scalar field systems ---see e.g. \cite{MacMinVal21}. 

\subsection*{Acknowledgements}
The problem addressed in this article was motivated by a conversation
with Helmut Friedrich at the Erwin Schr\"odinger Institute (ESI)
during the Workshop: Geometric Transport Equations in General
Relativity in 2017.  We have also profited from discussions with the
members of the online \emph{Conformal/spinorial workshop} held weekly
during the months of lockdown in 2020/2021: Alfonso Garcia-Parrado,
David Hilditch, Alex Va\~n\'o-Vi\~nuales, Marica Minucci, Mariem Magdy
Ali Mohamed, Greg Taujanskas, Peng Zhao, Tongtong Hu and Marco Luna.
During part of this project, EG was supported via the
European Union (through the PO FEDER-FSE Bourgogne 2014/2020 program)
and the EIPHI Graduate School (contract ANR-17-EURE-0002) as part of
the ISA 2019 project.
EG also acknowledges support by the FCT (Portugal)
2020.03845.CEECIND during the last stages of this work.


\begin{thebibliography}{10}

\bibitem{MagVal21}
M.~M. {Ali Mohammed} \& J.~A. {Valiente Kroon},
\newblock {\em A comparison of Ashtekar's and Friedrich's formalisms of spatial
  infinity},
\newblock in {\tt arXiv:2103.02389[gr-qc] }, to appear in
Class. Quantum Grav.  (2021).

\bibitem{Bei91b}
R.~Beig,
\newblock {\em Conformal properties of static spacetimes},
\newblock Class. Quantum Grav. {\bf 8}, 263 (1991).

\bibitem{ChrDel03}
P.~T. Chru\'{s}ciel \& E.~Delay,
\newblock {\em On mapping properties of the general relativistic constraint
  operator in weighted function spaces, with applications},
\newblock Mem. Soc. Math. France {\bf 94}, 1 (2003).

\bibitem{CorSch06}
J.~Corvino \& R.~Schoen,
\newblock {\em On the asymptotics for the Einstein Constraint Vacuum
  Equations},
\newblock J. Diff. Geom. {\bf 73}, 185 (2006).

\bibitem{Fri88}
H.~Friedrich,
\newblock {\em On static and radiative space-times},
\newblock Comm. Math. Phys. {\bf 119}, 51 (1988).

\bibitem{Fri91}
H.~Friedrich,
\newblock {\em {On the global existence and the asymptotic behaviour of
  solutions to the Einstein-Maxwell-Yang-Mills equations}},
\newblock J. Diff. Geom. {\bf 34}, 275 (1991).

\bibitem{Fri98a}
H.~Friedrich,
\newblock {\em Gravitational fields near space-like and null infinity},
\newblock J. Geom. Phys. {\bf 24}, 83 (1998).

\bibitem{Fri03b}
H.~Friedrich,
\newblock {\em Spin-2 fields on Minkowski space near space-like and null
  infinity},
\newblock Class. Quantum Grav. {\bf 20}, 101 (2003).

\bibitem{Fri04}
H.~Friedrich,
\newblock {\em Smoothness at null infinity and the structure of initial data},
\newblock in {\em 50 years of the Cauchy problem in general relativity}, edited
  by P.~T. Chru\'{s}ciel \& H.~Friedrich, Birkhausser, 2004.

\bibitem{Fri13}
H.~Friedrich,
\newblock {\em Conformal structure of static vacuum data},
\newblock Comm. Math. Phys. {\bf 321}, 419 (2013).

\bibitem{Fri18}
H.~Friedrich,
\newblock {\em Peeling or not peeling ---is that the question?},
\newblock Class. Quantum Grav. {\bf 35}, 083001 (2018).

\bibitem{FriKan00}
H.~Friedrich \& J.~K\'{a}nn\'{a}r,
\newblock {\em Bondi-type systems near space-like infinity and the calculation
  of the {N}{P}-constants},
\newblock J. Math. Phys. {\bf 41}, 2195 (2000).

\bibitem{GarMar12}
A.~Garc\'{\i}a-Parrado \& J.~M. Mart\'{\i}n-Garc\'{\i}a,
\newblock {\em {Spinors: a Mathematica package for doing spinor calculus in
  General Relativity}},
\newblock Comp. Phys. Commun. {\bf 183}, 2214 (2012).

\bibitem{GasVal17a}
E.~Gasperin \& J.~A.~Valiente Kroon,
\newblock {\em {Polyhomogeneous expansions from time symmetric initial data}},
\newblock Class. Quant. Grav. {\bf 34}, 195007 (2017).

\bibitem{GasVal20}
E.~Gasperin \& J.~A.~Valiente Kroon,
\newblock {\em {Zero rest-mass fields and the Newman-Penrose constants on flat
  space}},
\newblock J. Math. Phys. {\bf 61}, 122503 (2020).

\bibitem{KopVal21}
J.~Kopi{\'n}ski \& J.~A.~Valiente Kroon,
\newblock {\em New spinorial approach to mass inequalities for black holes in
  general relativity},
\newblock Physical Review D {\bf 103}, 024057 (Jan 2021).

\bibitem{MacMinVal21}
R. P. Macedo, M. Minucci \&  J.~A.~Valiente Kroon,
\newblock{\em {The Maxwell-scalar field near spatial infinity}},
\newblock in preparation.

\bibitem{Pae14}
T.-T. Paetz,
\newblock {\em {Killing Initial Data on spacelike conformal boundaries}},
\newblock J. Geom. Phys. {\bf 106}, 51--69 (2016).

\bibitem{PenRin84}
R.~Penrose \& W.~Rindler,
\newblock {\em Spinors and space-time. {V}olume 1. {T}wo-spinor calculus and
  relativistic fields},
\newblock Cambridge University Press, 1984.

\bibitem{PenRin86}
R.~Penrose \& W.~Rindler,
\newblock {\em Spinors and space-time. {V}olume 2. {S}pinor and twistor methods
  in space-time geometry},
\newblock Cambridge University Press, 1986.

\bibitem{Som80}
P.~Sommers,
\newblock {\em Space spinors},
\newblock J. Math. Phys. {\bf 21}, 2567 (1980).

\bibitem{Ste91}
J.~Stewart,
\newblock {\em Advanced general relativity},
\newblock Cambridge University Press, 1991.

\bibitem{Sza94b}
L.~B. Szabados,
\newblock {\em Two-dimensional Sen connections and quasi-local
  energy-momentum},
\newblock Class. Quantum Grav. {\bf 11}, 1847 (1994).

\bibitem{Sza94a}
L.~B. Szabados,
\newblock {\em Two-dimensional Sen connections in general relativity},
\newblock Class. Quantum Grav. {\bf 11}, 1833 (1994).

\bibitem{Val00}
J.~A. Valiente~Kroon,
\newblock {\em On the existence and convergence of polyhomogeneous expansions
  of zero-rest-mass fields},
\newblock Class. Quantum Grav. {\bf 17}, 4365 (2000).

\bibitem{Val03a}
J.~A. Valiente~Kroon,
\newblock {\em Polyhomogeneous expansions close to null and spatial infinity},
\newblock in {\em The Conformal Structure of Spacetimes: Geometry, Numerics,
  Analysis}, edited by J.~Frauendiener \& H.~Friedrich, Lecture Notes in
  Physics, page 135, Springer, 2002.

\bibitem{Val04d}
J.~A. Valiente~Kroon,
\newblock {\em Does asymptotic simplicity allow for radiation near spatial
  infinity?},
\newblock Comm. Math. Phys. {\bf 251}, 211 (2004).

\bibitem{Val04a}
J.~A. Valiente~Kroon,
\newblock {\em A new class of obstructions to the smoothness of null infinity},
\newblock Comm. Math. Phys. {\bf 244}, 133 (2004).

\bibitem{Val07a}
J.~A. Valiente~Kroon,
\newblock {\em Asymptotic properties of the development of conformally flat
  data near spatial infinity},
\newblock Class. Quantum Grav. {\bf 24}, 3037 (2007).

\bibitem{Val09a}
J.~A. Valiente~Kroon,
\newblock {\em {Estimates for the Maxwell field near the spatial and null
  infinity of the Schwarzschild spacetime}},
\newblock J. Hyp. Diff. Eqns. {\bf 6}, 229 (2009).

\bibitem{Val11}
J.~A. {Valiente Kroon},
\newblock {\em Asymptotic simplicity and static data}, {\bf 13}, 363 (2011).

\bibitem{CFEBook}
J.~A. {Valiente Kroon},
\newblock {\em Conformal methods in General Relativity},
\newblock Cambridge University Press, 2016.

\bibitem{Val10}
J.~Valiente~Kroon,
\newblock {\em A rigidity property of asymptotically simple spacetimes arising
  from conformally flat data},
\newblock Comm. Math. Phys. {\bf 298}, 673 (2010).

\bibitem{Wei90a}
G.~Weinstein,
\newblock {\em On rotating black holes in equilibrium in general relativity},
\newblock Communications on Pure and Applied Mathematics {\bf 43}(7), 903
  (1990).

\end{thebibliography}
\end{document}